\documentclass[letterpaper]{article}
\usepackage{amsmath}
\usepackage{dsfont}
\usepackage{aaai}
\usepackage{times}
\usepackage{helvet}
\usepackage{algorithm}
\usepackage{algpseudocode}
\usepackage{courier}
\usepackage{setspace}
\usepackage{amsmath}
 \usepackage{color, colortbl}
\usepackage{graphicx}
\usepackage{multicol}
\usepackage{balance}
\usepackage{amsmath}
\usepackage{amsfonts}

\newcommand{\modelch}[1]{\textsc{{#1}}\noindent\normalsize}
 \newcommand{\eqnref}[1]{Eq.~(\ref{#1})}

\newcommand{\VIP}{\textsc{Vip}}

\newcommand{\commentout}[1]{%
}

\newcommand{\figref}[1]{Figure~\ref{#1}}
\newcommand{\tabref}[1]{Table~\ref{#1}}

\newcommand{\remove}[1]{}
\definecolor{LightCyan}{rgb}{0.88,1,1}
\definecolor{Gray}{gray}{0.7}

\begin{document}
 \bibliographystyle{aaai}
\title{User Effort and Network Structure Mediate Access to Information in Networks}
\author{
Jeon-Hyung Kang \\
USC Information Sciences Institute\\
4676 Admiralty Way \\
Marina del Rey, CA 90292, USA \\
\texttt{jeonhyuk@usc.edu} \\
\And
Kristina Lerman \\
USC Information Sciences Institute\\
4676 Admiralty Way \\
Marina del Rey, CA 90292, USA \\
\texttt{lerman@isi.edu}
}
\maketitle

\begin{abstract}
Individuals' access to information in a social network depends on its distributed and where in the network individuals position themselves. 
However, individuals have limited capacity to manage their social connections and process information. 
In this work, we study how this limited capacity and network structure interact to affect the diversity of information social media users receive. 
Previous studies of the role of networks in information access were limited in their ability to measure the diversity of information. We address this problem by learning the topics of interest to social media users by observing messages they share online with their followers. We present a probabilistic model that incorporates human cognitive constraints in a generative model of information sharing. We then use the topics learned by the model to measure the diversity of information users receive from their social media contacts.
We confirm that users in structurally diverse network positions, which bridge otherwise disconnected regions of the follower graph, are exposed to more diverse information. In addition, we identify user effort as an important variable that mediates access to diverse information in social media.
Users who invest more effort into their activity on the site not only place themselves in more structurally diverse positions within the network than the less engaged users, but they also receive more diverse information when located in similar network positions.
These findings indicate that the relationship between network structure and access to information in networks is more nuanced than previously thought.
 
\end{abstract}
\section{Introduction}
 People use their social contacts to gain access to information in social networks~\cite{granovetter1973,Burt04}, which they can then leverage for personal advantage. However information in social networks is non-uniformly distributed, leading sociologists to explore the relationship between an individual's \emph{network position} and the novelty and diversity of information she receives through her social contacts.
Studies of social and organizational networks identified the importance of so-called brokerage positions, which link individuals to otherwise unconnected people~\cite{granovetter1973,Burt95,burt2005brokerage,Aral11}. By spanning distinct communities, brokerage positions expose individuals to novel and diverse information, which leads to new job prospects~\cite{granovetter1973} and higher compensation~\cite{Burt95,Burt04}.
However, the links that connect individuals in brokerage positions to the rest of the network,  generally represent weaker relationships (i.e., acquaintances rather than close friends)~\cite{granovetter1973,onnela2007structure}. The less frequent interactions along these ``weak'' links limit the amount of  information flowing to individuals~\cite{Aral11}. 
Thus, those who are able, and willing, to invest greater effort in social interactions, will manage more connections thereby increasing the volume of information they receive through those links~\cite{aral2012anatomy,Miritello13}.
Specifically, Aral \& Van Alstyne~\cite{Aral11} showed that individuals can increase the diversity and novelty of information they receive via email either by placing themselves in brokerage positions, or by communicating more frequently with their social contacts.

In contrast to email and phone interactions, where information is exchanged between a pair of social contacts, social media users broadcast information to all their contacts. Bakshy et al.~\cite{bakshy2012role} showed that weak links collectively deliver more novel information to Facebook users, even though they interact infrequently with these contacts. These findings suggest that an easy way for social media users to increase their access to diverse information is by creating more links, e.g., by following other users.
However, cognitive  (and temporal) constraints limit an individual's capacity to manage social interactions~\cite{Dunbar,goncalves2011modeling,Miritello13} and process the information they receive~\cite{Weng:2012dd,Hodas12socialcom}.
In addition, social media users vary greatly in the effort they expend engaging with the site, leading to a large variation in user activity, as measured by the number of messages posted on the site~\cite{Wilkinson08}. The impact of this variation on the information individuals receive and their position in the network is not known. Do users who are able (or at least willing) to be more active on the site receive more diverse information? Do they curate their social links so as to move themselves into network positions that provide more diverse information?

In this work, we   use data from the microblogging site Twitter to study the interplay between network structure, the effort Twitter users are willing to invest in engaging with the site, and the diversity of information they receive from their contacts. Previous studies of the role of networks in individual's access to information were limited in their ability to measure the diversity of information, using bag-of-words~\cite{Aral11} or predefined categories~\cite{kang2013structural} for this task. In this work, we learn topics of interest to social media users from the messages they share with their followers. We present a probabilistic topic model that incorporates human cognitive constraints in a generative model of information sharing and evaluate the model on the task of predicting the messages users retweet. We demonstrate that our model has competitive performance, and unlike other models, it produces descriptions of topics.

We use learned topics to measure the diversity of information users receive from their contacts. This enables us to study the factors that affect the diversity of information in networks.
Our findings indicate that the relationship between network structure and access to information is more nuanced than previously thought.
First, users cannot increase the diversity of the information they receive by increasing the number of their contacts.  Second, we confirm that users in structurally diverse network positions, which bridge otherwise disconnected regions of the follower graph, are exposed to more diverse topics via their contacts than users in less structurally diverse positions.
However, we demonstrate that user effort is an important variable mediating access to information in networks. Active users who post more messages on Twitter receive more diverse information even when they are in structurally similar positions to the less active users. This suggests that users who are willing (or able) to engage more on Twitter curate their contacts so as to increase the diversity of the information they receive. Since effort is a useful proxy for individual's cognitive capacity for (or at least the willingness to invest the time in) processing information in social networks~\cite{Miritello13b}, our work suggests that cognitive factors interact in non-trivial ways with network structure to define access to information in social networks.

\section{Description of Data}
\label{sec:data}
Twitter is an online social networking and microblogging service that allows users to follow the activity of others to see the messages they posted or retweeted recently. When a user posts or retweets a message, it is broadcast to all her followers, who are then able to see it in their own streams. Twitter offers an Application Programming Interface (API) for data collection.
We used two data sets collected in the past from Twitter. The 2012 data set~\cite{Kang15sbp} contains tweets including a \emph{URL} to monitor information spread over the social network from Nov 2011 to Jul 2012.
They start by monitoring potential seed \emph{URL}s containing \emph{http://t.co} from the streaming APIs and collect all tweets containing them.
Since the total volume of tweets containing a \emph{URL} is very large, they focus on broadly shared \emph{URL}s.
They selected as seeds the \emph{URL}s that appeared more than once in five days from its initial appearance in the streaming APIs based on the heuristic that the \emph{URL}s that have been appeared more often in the streaming APIs will be more popular on Twitter.
They collected the entire history of these seed \emph{URL}s
until there were no more tweets containing them within five days from their last appearance in the Twitter REST APIs.
 This yielded 12.5M tweets with 9.5M users.

The 2014 data set contains the tweets from 5600 initial seed users~\cite{Smith13socialcom} and their friends from Mar 2014 to Oct 2014. Starting with 5,600 initial seed users, they collected all their friends and at least first 200 tweets from their time line. 
The data set includes 23.8 M tweets from 1.9M users with 17.8M social network links.

\section{Probabilistic Model of User Topics}
\label{sec:xxvip}

We use a probabilistic model to learn users' topics of interest from the messages they share in social media. What information users share, and which messages shared by friends they decide to spread to their followers, depends on a number of factors, such as virality of information being shared, users' tastes, and their followers' tastes.
To understand information sharing in social networks, social recommendation models~\cite{ma2008sorec,WangB11,Kang13sbp} were used to represent users' interests and items they share by $k$-dimensional topic vectors.
Once these hidden topic vectors are learned from user's item adoption (i.e., retweeting) history, it is possible to calculate the \emph{personal relevance} of a new item to the user.

\begin{figure}[t]
\begin{center}
\includegraphics[width=0.75\linewidth]{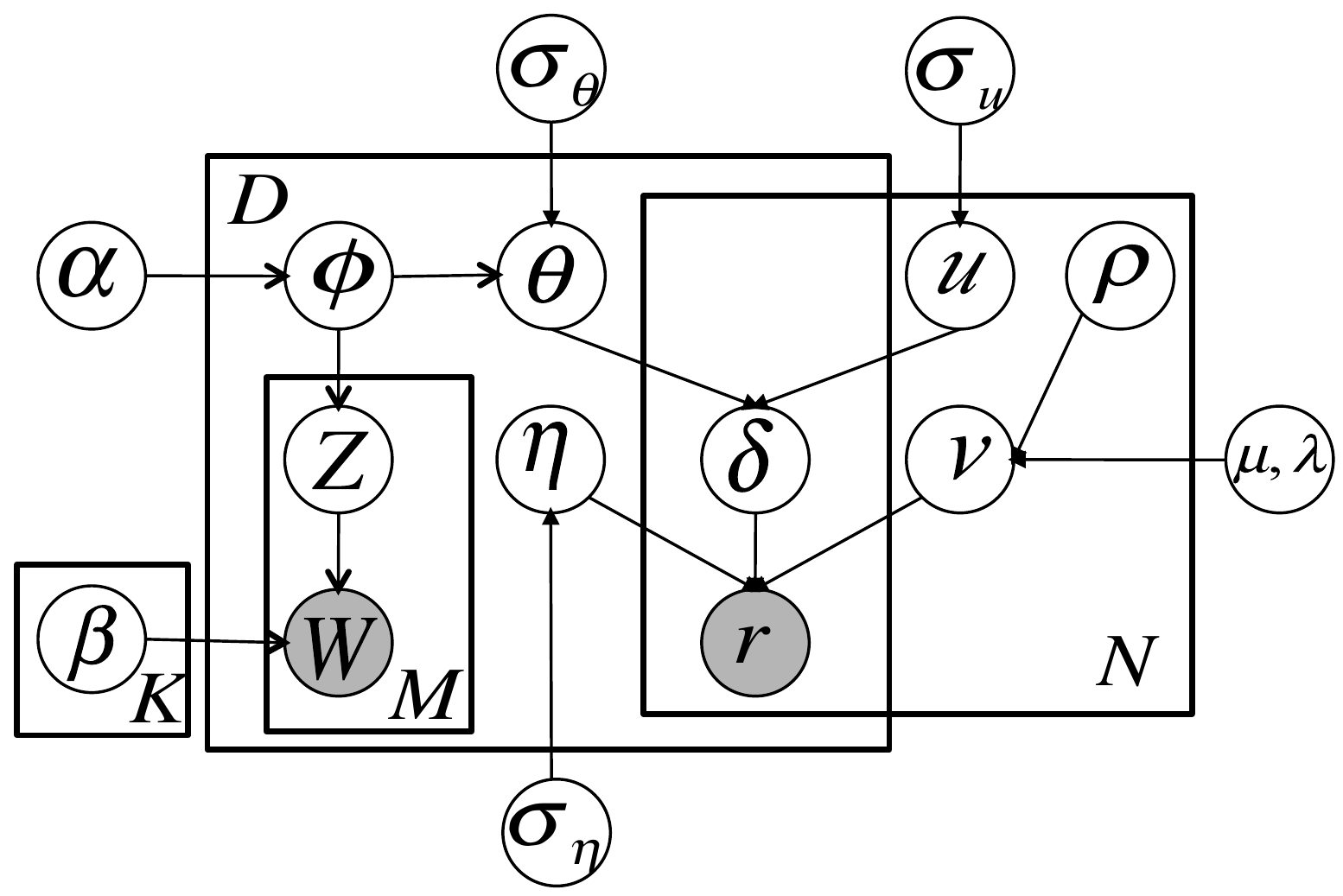}
\end{center}
\caption{Our model  with user topic ($u$) and item topic ($\theta$) profiles, item's personal relevance ($\delta$) and visibility to user ($v$), item fitness ($\eta$), expected number of new posts user received ($\rho$) and item adoption ($r$). Topic model part has the topic distribution ($\phi$) of an item and a distribution($\beta$) over words from a vocabulary of size ${M}$. $N$ is the number of users, and ${D}$ is the number of items.}
\label{fig:modelDiagram}
\end{figure}

We proposed {\VIP}~\cite{Kang15sbp}, a model that captures the three basic ingredients of information spread in social media: item's visibility ($v$) to a user, its fitness or virality ($\eta$),  and its (personal) relevance ($\delta$) to the user.
While the model improves on previous models, it applies normal distribution assumptions on modeling binary responses, uses full user-item adoption matrix, and provides no descriptions on the learned latent topic space. In this paper, we model binary responses (adopted vs unadopted items) of social media users with multinomial logic model. Stochastic optimization allows us to learn from randomly sampled negative (not adopted) and positive (adopted) dyads without overfitting to the positive ones. Our stochastic inference algorithm handles many user-item dyads and can be distributed for efficient computation. Furthermore, with the help of a probabilistic topic model, we can provide an interpretable low-dimensional representation of information.
Figure~\ref{fig:modelDiagram} graphically represents our model.

\paragraph{Item visibility}
When a user's message stream is delivered as a list of items, the process of item discovery is biased by the position of each item in the list. A user is more likely to see items near the top of the list than those deeper in the stream~\cite{Lerman14plosone}. Hence, items in top stream positions have higher visibility. Since we do not know an item's exact position, we estimate it as the average visibility of items to user $i$ as follows:
\begin{equation}
\begin{aligned}
& v_{i} \sim \sum_{L} \left( \mathbf{G}(1/(1+\rho_i),L) (1-\mathbf{IG}(\mu, \lambda, L)) \right) \\
 \end{aligned}
  \label{eq:visibile}
\end{equation}
The first factor gives the probability that user $i$ discovers an item depending on the number of items in her stream.
The greater the number of new messages user receives between visits to the site, the less likely the user is to view any specific item. Thus, average visibility depends on the frequency the user visits the site and the rate of posts received.
This competition between the rates friends post new messages to the user's stream and the rate user visits the stream to read the messages modeled by a geometric distribution with success probability $p=1/(1+\rho_i)$: $\mathbf{G} = (1-p)^L p$. The ratio $\rho_i$ of these rates gives the expected number of new messages in a user's stream.
The second factor of gives the probability that user $i$ will navigate to at least $(L+1)$-th position in the stream to view the item.  This is estimated by  the upper cumulative distribution of an inverse gaussian $\mathbf{IG}$ with mean $\mu$ and shape parameter $\lambda$ and variance $\mu^3/\lambda$:
\begin{equation}
\begin{aligned}
& \exp \left(\frac{-\lambda  (L-\mu )^2}{2\mu ^2L} \right) \left[\frac{\lambda }{2\pi L^3} \right]^{(1/2) }.
 \end{aligned}
\end{equation}

\paragraph{Item virality}
Social media users adopt items even if they had not earlier demonstrated a sustained interest in their topics. This is often the case with viral, general-interest items, such as breaking news or celebrity gossip. Thus, we use ``virality'' to represent item's propensity to spread on exposure.
\begin{equation}
\begin{aligned}
 & \eta_j  \sim   \mathcal{N}  (0, \sigma_\eta^{2})  
  \end{aligned}
\end{equation}

\paragraph{Item relevance}
We calculate personal relevance of an item $j$ to user $i$ as:
\begin{equation}
\begin{aligned}
& \delta_{ij} \sim g_\delta({u_i^T \theta_{j}})\\
 \end{aligned}
\end{equation}
where symbol $T$ refers to the transpose operation, $u_i$ represents the topic profile of user $i$, $\theta_j$ represents the topic profile of item $j$ and $g_\delta$ is linear function for simplicity.
\begin{equation}
\begin{aligned}
& u_{i} \sim \mathcal{N}(0,  \sigma_{u}^{2}I_{K})\\
& \theta_{j} \sim \mathcal{N}(0,  \sigma_{\theta}^{2}I_{K})\\
 \end{aligned}
\end{equation}
where $K$ is the number of topics.

We use a widely known text mining algorithm Latent Dirichlet Allocation (LDA)~\cite{blei2003latent}, which analyzes the co-occurrence of the words in documents, to learn the hidden topics representing the documents. In our case, LDA captures the item's topic distribution $\phi$, which is represented as $K$ dimensional vector in the recommendation model. The topic distribution of each document ($\phi_{d_j}$) is viewed as a mixture of multiple topics, with each topic ($\beta_{k}$) as a distribution over words. In our setting, the corpus $D$ is a collection of tweet text of the tweet posts. The likelihood of $D$ is computed by multiplying over all documents and all words in each document as follows:

\begin{equation}
\begin{aligned}
p(D | \beta_, \phi, z )  = \prod_{d_j \in D} \prod_{w \in d_j} {\phi_{d_j,z_{w}} \beta_{z_{w},w}}
 \end{aligned}
\end{equation}
where $z_w$ is assigned topic index for each word $w$ in the document $d_j$, $\phi_{d_j,z_{w}}$ is the likelihood of topics $z_{w}$ for the document $d_j$ and $\beta_{z_{w},w}$ is the likelihood of choosing specific word $w$ for the topic $z_{w}$.

The generative process for item adoption through a social stream can be formalized as follows:
\begin{tabbing}
For \=each user $i$ \\
\>  Generate \= $u_{i} \sim \mathcal{N}(0,\sigma_{u}^{2}I_{K}$)\\
\>  Generate \= $v_{i} \sim \sum_{L} \left( \mathbf{G}(1/(1+\rho_i),l) (1-\mathbf{IG}(\mu, \lambda,l)) \right)$ \\
For \=each item j \\
\>  Generate \= $\eta_{j} \sim \mathcal{N}(0,\sigma_{\eta}^{2}$)\\
\>  Generate \= $\phi_{j} \sim Dirichlet({\alpha}$) \\
\>  Generate \= $\epsilon_{j} \sim \mathcal{N}(0,\sigma_{\theta}^{2}I_{K}$) and set \= $\theta_{j} = \epsilon_{j} + \phi_{j}$\\
\>\;\;\;\;\;\;For each word $w_{jm}$ \\
\>\>Generate topic assignment $z_{jm} \sim Mult ( \phi_j$)\\
\>\> Generate word $w_{jm} \sim Mult ( \beta_{z_{jm} }$)\\

 For \=each  user $i$\\
\> For each item $j$ on the news feed\\
\>\;\;\;\;\;\;Generate the adoption $r_{ij} \sim p(I({r_{ij}})  | u_i, v, \theta, \eta, O_i ) $ 
\end{tabbing} 
 Lack of adoption by user $i$ of item $j$ ($r_{ij}=0$) can be interpreted in two ways: either the user saw the item but did not like it, or the user did not see the item but may have liked it had she seen it.
While other models partly account for the lack of knowledge about non-adoptions using smoothing~\cite{WangB11,Kang13aaai},  we properly model visibility of items to users.

We model the user-item adoption with Softmax function, which makes the values of the $K$ dimensional vectors in [0-1] range. The equation is as follows:
\begin{equation}
\begin{aligned}
p(I({r_{ij}})  | u_i, v, \theta, \eta, O_i )  = \frac{\exp(v_i g_r( \delta_{ij} + \eta_j))}{\sum_{l \in O_i} \exp(v_i g_r( \delta_{il} + \eta_l))}
 \end{aligned}
\end{equation}
\noindent where $I({r_{ij}})$ is the indicator function, $ I({r_{ij}})$ = 1 when user $i$ adopted item $j$ and 0 otherwise, and $O_i$ is the observed items by user $i$. We define $g_r$ as linear functions for simplicity.

The main objective function is:
 \begin{equation}
 \label{equ:min}
\begin{aligned}
\ell =
& -\frac{1}{2 \sigma_u^{2}} \sum_i^N u_i^T u_i -\frac{1}{2 \sigma_\eta^2} \sum_j^D {\eta_{j}}^T {\eta_{j}} \\
& -\frac{1}{2 \sigma_\theta^2} \sum_j^D { (\theta_j-\phi_j)^T  (\theta_j-\phi_j)}\\
& +\sum_i^N \log\left(\sum_l^L (1/\rho_i+1)(\rho_i/\rho_i+1)^l (1-\mathbf{IG}(\mu,\lambda,l)  \right)\\
& - \sum_{i}^N\sum_{j}^{D} \left(\log ( \sum_{l \in O_i} \exp( v_i({ \delta_{il}+\eta_l})) ) -  v_i({ \delta_{ij}+\eta_j}) \right)\\
 \end{aligned}
\end{equation}
The last term of the equation minimizes the error between the binary rating and the predicted rating. The second line of the equation minimizes the error between the topics that explain the recommendation and the content. The importance between these two components can be controlled with $\sigma_\theta$.
MAP estimation is equivalent to maximizing the complete log likelihood ($\ell$) of $U$, $V$, $\theta$, $\eta$, $\phi$ and ${r}$ given $\sigma_u$, $\sigma_\theta$, $\sigma_\eta$, $\mu$, $\lambda$ and $\rho$.

\subsection{Model Learning}
\label{sec:learningmodel}
To optimize \eqnref{equ:min}, we develop a stochastic gradient descent algorithm. Given a current estimate, we take the gradient of \eqnref{equ:min} with respect to $u_i$, $ \theta_j$, and $\eta_j$ and iteratively optimize the parameters \{$u_i, \theta_j, \eta_j$\}. Derived update equations are: 

\begin{algorithm}
\caption{Stochastic Optimization }
\begin{algorithmic}
    \State Initialize model parameter ${U, V, \theta, \eta, \phi, \nabla}$
    \For  {$t=1$ to $T$}
		\For  {$u$ in $U$}	
			\State Choose random $|r_i|$ mini batch $S_i$ from $D$-$r_i$
			\State Generate $O_i = r_i \cup S_i $
			\For  {$j$ in $O_i$}	
				\State $u_i \leftarrow  u_i - \mu\ [  v_j  \theta_j \nabla   + \frac{1}{2| r_{i }|\sigma_u^{2} }  u_i ]$
	    	   		\State $\theta_j \leftarrow \theta_j - \mu \ [   v_i u_i \nabla   + \frac{1}{2|r_{\cdot j}|\sigma_\theta^{2}}   (\theta_j - \phi_j)]$		
				\State $\eta_j \leftarrow  \eta_j - \mu\ [  v_i \nabla   + \frac{1}{2|r_{\cdot j}| \sigma_\eta^2}  \eta_j ]$
			\EndFor
		\EndFor
    \EndFor
\end{algorithmic}\label{alg}
\end{algorithm}
\noindent where $|r_{i }|$ is the number of items adopted by user $i$ and $|r_{\cdot j}|$ is the number of users who adopted item $j$. We generate a set of observed items $O_i$ by adding randomly sampled $|r_i|$ number of items from the unadopted set ($D$-$r_i$) and incrementally learning from the unadopted and adopted item set of each user.
We use the learning rate $\mu$ with discount by a factor of 0.9 in each iteration \cite{koren2009matrix}.

The equation for gradient ($\nabla$) is as follows:
\begin{equation}
\begin{aligned}
\nabla = &   \frac{\exp(v_i g_r( \delta_{ij} + \eta_j))}{\sum_{l \in O_i} \exp(v_i g_r( \delta_{il} + \eta_l))} - I({r_{ij}}).
 \end{aligned}
\end{equation}
 The proposed recommendation model can be updated incrementally to model dynamic user adoptions in real time. It is also computationally efficient since it can be distributed by decomposing the data set over multiple computers.

\begin{table}
  \centering
\caption{Model parameters used in this study.}
\scalebox{0.9}{
\begin{tabular}{|c|c|}
  \hline
\cellcolor{Gray}{Parameters}	&	\cellcolor{Gray}{Value}	\\
\hline
number of topics	&	K =100	\\
\hline
user topic profile 	&	 $\sigma_u^{2} $=$10^{4}$	\\
item topic profile	&	 $\sigma_\theta^{2} $=$10^{4}$	\\
item fitness 	&	 $\sigma_\eta^{2} $=$10$	\\
\hline
law of surfing 	&	$\mu=14.0$ \\
  	&	$\lambda=14.0$ \\
	\hline
views per post 	&	38	\\
	\hline
	typical posting rates   	&	 1.4 \\
\hline
\end{tabular}
}
\label{tbl:modelparam}
\end{table}

\subsection{Model Selection}
\label{sec:ms}
We use the same ``law of surfing'' parameters, $\mu=14.0$ and $\lambda=14.0$, as \cite{Kang15sbp,Hogg13socialcom,Hogg12epj} did in their study of social media. The expected number of new posts including a URL user $i$ received, $\rho_i$, is computed by $rate^{(url\ posts\ received)}_i/rate^{(visits)}_i$. The rate $rate^{(posts\ received)}_i$ is proportional to the  number of friends ($N_{frd(i)}$) $i$ follows and their average posting frequency. To estimate posting frequency of all users, we use the typical URL posting rates of users from our data: $rate^{(posts\ received)}_i=1.4*N_{frd(i)}$.
We estimate user $i$'s visiting rate ($rate^{(visits)}_i$) using the number of posts of user $i$ ($N_{posts(i)}$).
 \cite{Hogg13socialcom} estimated that average number of visits per post was 38 (2014 data set) for Twitter users. Also, since around 20\% of tweets include a URL~\cite{chaudhry2012trends}, the posting rate of user $i$ becomes $rate^{(visits)}_i = 7.6*N_{posts(i)}$ (2012 data set).

For the model hyper-parameters, we vary the parameters $K\in $\{10, 30, 50, 100, 200\}, and  \{$\lambda_u, \lambda_\theta \} \in $\{$10^{-4}$, $10^{-3}$,..., $10^4$\} by using grid search on validation set. Throughout this paper, we set parameters $K=100$, $\lambda_u  = 0.01$, $\lambda_\theta  = 0.001$, both for PMF and CTF that performed the best for PMF. For the fitness parameter of VIP~\cite{Kang15sbp} and the proposed model, we vary $\sigma_\eta^2 \in $ \{$10^{-4}$, $10^{-3}$,..., $10^4$\}, while we fix other parameters: $\sigma_\theta^2= 10^4$ and $\sigma_u^2=10^4$. In this paper, we set $\sigma_\eta^2=10$.

\subsection{Model Evaluation}
We evaluate the proposed model by using it to predict which items users will adopt. For this task, user $i$'s adoption of item $j$ shared by a friend is obtained by point estimation with optimal variables \{$\theta^{*}$, $u^{*}$, $v^{*}$, $\eta^{*}$\}:
\begin{equation}
\begin{aligned}
\mathbb{E}[r_{ij}|\mathcal{D}] \approx&
     \mathbb{E}{[v_{i}|\mathcal{D}]}^T  (\mathbb{E}[\delta_{ij}|\mathcal{D}] +\mathbb{E}[\eta_{j}|\mathcal{D}] )\\
r_{ij}^{*} \approx& {v_{i}^{*}} ( {u_{i}^{*}}^T{\theta_{j}^{*}}+{\eta_{j}^{*}})
 \end{aligned}
\end{equation}
where $\mathcal{D}$ is the training data. The adoption probability is decided by user visibility $v^{*}_{i}$, user topic profile  $u^{*}_i$, item topic profile $\theta^{*}_j$, and item fitness  $\eta^{*}_j$.

To evaluate the performance, we use precision (P), recall (R) and normalized discounted cumulative gain (nDCG) for top-x recommended posts.
\begin{description}
\item[P@$x$ ] computes the fraction of items that are adopted by each user in top-$x$ items in the list. We average the precision@$x$ of all users.
\item[R@$x$ ] computes the fraction of adopted items that are successfully discovered in top-$x$ ranked list out of all adopted items by each user. We average the recall@$x$ of all users.
\item[nDCG@$x$ ] computes the weighted score of adopted items based on the position in the top-$x$ list. It penalizes adopted items in the bottom of the top-$x$ list. We average the nDCG@$x$ of all users.
\end{description}

\begin{table}
  \centering
\caption{Overall prediction performance comparison using Precision@$x$ (P@$x$), Recall@$x$ (R@$x$), normalized DCG@$x$ (nDCG@$x$) on Twitter dataset.}
{
\begin{tabular}{|c|c|c|c|c|c|}
\hline 	
\cellcolor{Gray}{Model}	&	\cellcolor{Gray}{\small{Text}}	&	\cellcolor{Gray}{\small{P@$10$}	}	&	 \cellcolor{Gray}{\small{R@$10$}}	&	\cellcolor{Gray}{\small{nDCG@$10$}}		\\
	\hline
Random	&	No	&	0.0483	&	0.3738	&	0.2410	\\
\hline
Fitness	&	No &0.0798	&	0.5924	&	0.3630	\\
\hline
Relevance	&	No		&	0.0647	&	0.4383	&	0.3170		\\		
\hline
{\VIP} 	&	No		&	0.0984	& 0.6446	& 0.4205	\\
\hline
Softmax-CTR&	Yes	  	&	0.1047	&	0.6105	&	0.4123	\\
\hline
\cellcolor{LightCyan}{Our Model} 	&	\cellcolor{LightCyan}{Yes}  	&	\cellcolor{LightCyan}{0.1138}	&	 \cellcolor{LightCyan}{0.7022}	&	\cellcolor{LightCyan}{0.4619}	\\
\hline
\end{tabular}
}
\label{tbl:overallcomparison}
\end{table}

We divide each user's adopted items into five folds and construct the training set and the test set. We use five-fold cross validation and compare performance of the proposed model to five baseline models: \modelch{Random}, \modelch{Fitness}, \modelch{Relevance}, \modelch{\VIP}, \modelch{CTR}. The \modelch{Random} baseline chooses items at random from among the items in user $i$'s stream, i.e., items adopted by $i$'s friends. The baseline \modelch{Fitness} uses item fitness values ($\eta$) learned by \modelch{\VIP} to recommend $k$ highest fitness items. The baseline \modelch{Relevance} bases its recommendations on user-topic and item-topic vectors learned by PMF. 
Collaborative Topic Regression (\modelch{CTR})~\cite{WangB11} was originally introduced to recommend scientific articles. It combines collaborative filtering (PMF) and probabilistic topic modeling (LDA). It captures two $K$-dimensional lower-rank user and item hidden variables from user-item adoption matrix and the content of the items. This model uses textual information and negative dyads, but unlike our method it uses $\ell_2$ function instead of a Softmax. Here for a fair comparison, we implemented a Softmax version. Based on our experiment Softmax-CTR outperformed original CTR due to the binary adoptions of social media.

\tabref{tbl:overallcomparison} shows the models' overall performance on the user--item adoption prediction task. In this paper, we set $x$=10 since recommending too many items is not realistic. From our experiments, we found that results are consistent with different number of $k$. While nDCG@$x$ uses the position of correct answer in the top-$x$ ranked list, it does not penalize for unadopted items or missing adopted items in the top-$x$ ranked list, therefore one has to consider the performance of all three metrics together. Intuitively a better model should have higher P@$x$, R@$x$, and nDCG@$x$.

 The experimental results show that the proposed model dramatically outperforms the random model with 135.61\% and 87.85\% respectively on precision and on recall. A comparison against the random model is important to uncover the complexity of the post-recommendation task.
 \modelch{Fitness} and \modelch{Relevance} models yield 62.21\% and 33.95\% improvement over the random model in terms of precision, and 58.48\% and 17.25\% in terms of recall respectively.
 The gain of \modelch{\VIP} over \modelch{Relevance}  is 52.08\% on precision and 47.06\% on recall, while the one of \modelch{CTR} over \modelch{Relevance}  is 61.82\% on precision and 39.28\% on recall.  This shows that accounting for cognitive biases dramatically improves predictability of user item adoptions in social media as much as accounting for text description of items alone.
 Among all models, the proposed model yields best performance, showing that modeling text, as well as visibility,  is critical in social media recommendation.

\section{Information Access in Networks}
\label{sec:analysis}
We use the topics learned by the proposed model to study how information is distributed in a network and what users can do to increase the diversity of information they receive from their social media friends.
In order to use the messages users posted, in addition to friends' messages they retweeted, we changed the model by assigning visibility equal to one to each original message user posted.

\subsection{Definition of Variables}
Following~\cite{Aral11,aral2012anatomy} we define a set of variables we use to characterize users, their network position, and information diversity.

\begin{table}
\center
\begin{tabular}{|l|l|}
\hline
\cellcolor{Gray}{\textbf{Var.}} &\cellcolor{Gray}{\textbf{   Description }  } \\ \hline
$S_i$ & number of active friends \\ 
$ND_i$ & network diversity \\  
$O_i$ & avg. vol. of outgoing info. (\# tweets/day) \\ 
$u_i$ & user-topic vector.  ($k$-dimensional vector)\\
$FTD_i$ & friend topic diversity \\ \hline 
\end{tabular}
\center
\caption{Variables used in the study.}
\label{tbl:definitions}
\end{table}

\paragraph{Network size}
We define the network size $S_i$ of user $i$ as the number of friends from whom user $i$ received messages during a time period $\Delta t$, which we take to be the data collection period. We only consider active friends, i.e., friends who posted messages during $\Delta t$.  Network size is defined as
 \begin{equation}
\begin{aligned}
S_i=\sum_{l \in N^{frd}_i}{I(r_{l })} 
 \end{aligned}
\end{equation}
\noindent where $N^{frd}_i$ is the set of friends of user $i$ and the indicator function $I(r_{l })$ is one if and only if friend $l$ tweeted during the time period $\Delta t$ and zero otherwise.

\paragraph{Network diversity}
User's position in a network significantly impacts the diversity of received information. Position can be characterized by its structural diversity, which represents how many otherwise unconnected contacts user $i$ has.
We measure structural diversity of a network position using local clustering coefficient~\cite{watts1998small}, $C_i$, which quantifies how often user $i$'s contacts are linked  (regardless of the direction of the link):
 \begin{equation}
\begin{aligned}
{C_i= \frac{ 2 \times |\{ e_{jk}: j, k \in N^{frd}_i, e_{jk} \in E \} |} { S_i (S_i-1)} } 
 \end{aligned}
\end{equation}
\noindent The variable $e_{jk}=1$ if user $j$ follows user $k$ or vice versa; otherwise, $e_{jk}=0$. The total number of possible connections among contacts is $S_i (S_i-1)$. High clustering coefficient implies low network diversity, and vice versa. Therefore, we define network diversity of user $i$ as $ND_i=1-C_i$. Note that brokerage positions have high network diversity, while individuals in tightly-knit communities have are in positions with low network diversity.

\paragraph{User effort} 
Most social media sites, including Twitter, display items from friends as a chronologically ordered list, with the newest items at the top. A user scans the list and if she finds an item interesting, she may share it with her followers by retweeting it. She will continue scanning the list until she loses interest or distracted~\cite{Hodas12socialcom}.
It is difficult to quantify how much of the list a user processes, since the site does not provide this information. Instead, we use user \emph{activity} as a heuristic for the effort users are willing (or able) to invest in Twitter.
We measure user $i$'s activity by the average number of messages the user tweets and retweets per day: 
 \begin{equation}
\begin{aligned}
{O_{i}= \frac{ |r_{i}|}{ \Delta t}} 
 \end{aligned}
\end{equation}
\noindent where $ |r_{i}|$ is the number of tweets from user $i$. 

\paragraph{Friend topic diversity}
We measure the diversity of information user $i$ receives from friends by the the variance of friends' topic interests: when most of friends have distinct, non-overlapping, interests, topic diversity will be high, whereas when most of friends have similar topic interests it will be low.
We define friend topic diversity as the average pair-wise cosine distance of friends' topic interest vectors.
\begin{equation}
\begin{aligned}
FTD_{i}= \frac{ 2 \times \sum_{j \in N^{frd}_i} \sum_{k \in N^{frd}_i} (1-Cos(u_{j},u_{k}))}{ {  S_i (S_i-1)} } 
 \end{aligned}
\end{equation}

\subsection{Information and Network Structure}
\begin{table}
  \centering
\caption{Keywords associated with the top 10 topics of users in different positions within the network. Users are divided into two populations based on their network diversity ($ND$).}
{
\begin{tabular}{|c|c|c|} 
\hline 
\cellcolor{Gray}{\small{\#}}	&\cellcolor{Gray}{Users in a Low ND}	&	\cellcolor{Gray}{{Users in a High ND}}	 	\\
\hline

\small{1} 	&\small{lesson weight loss acoustic} 	&	\small{profession connect profile } \\
\tiny{} 	&\small{lose motive guitar flash gain}	&	\small{webdesign bigdata update} 	\\ 
\hline
\small{2} 	&\small{pet dog  animal adopt praise}  	&	\small{children parent surgery inch}  \\
\tiny{} 	&\small{cat rescue love mate relax } &	\small{anxiety obesity autism} 	\\ 
\hline
\small{3} 	&\small{read book review kindle} 	&	\small{united kingdom stadium}   \\ 
\tiny{} 	&\small{novel cover publish buddha}	&	\small{arena holland yankees} 	\\ 
\hline
\small{4} 	&\small{good happy hope morn} 	&	\small{prosecute labour governor}  \\ 
\tiny{} 	&\small{ birthday wish love like}	&	\small{palestinian nationwide peru} 	\\
\hline
\small{5} 	&\small{yoga workout exercise jump}	&	\small{ferguson pray  brooklyn}  \\ 
\tiny{} 	&\small{doctor fit body back diet} 	&	\small{documentary oakland } 	\\ 
\hline
\small{6} 	&\small{graphic japanese poetry }  	&	\small{art center science exhibit}    \\
\tiny{} 	&\small{manga cinema photo  }	&	\small{culture paper draw museum} 	\\ 
\hline
\small{7} 	&\small{oil kale gene napa sausage}  	&	\small{camera shoot timeline canon}    \\
\tiny{} 	&\small{wrap aspire coal trainer}	&	\small{len accent timeline possess} 	\\ 
\hline
\small{8} 	&\small{children parent common}  	&	\small{worldcup shout football}    \\
\tiny{} 	&\small{journey ready pack escape}	&	\small{soccer illinois  player sold  } 	\\ 
\hline
\small{9} 	&\small{home design studio site}  	&	\small{space mars nasa planner }    \\
\tiny{} 	&\small{interior built lawn layout}	&	\small{newton isaac modern} 	\\ 
\hline
\small{10} 	&\small{beauty summer city park}  	&	\small{free win get email gift }    \\ 
\tiny{} 	&\small{resort nation beach island}	&	\small{chance enter offer ticket} 	\\ 
\hline
\end{tabular}
}
\label{tbl:topicscomparison}
\end{table}

Information is not uniformly distributed in a network: users in brokerage positions are interested in systematically different topics than users within denser communities. To study user-topic distribution, we rank users according to network diversity ($ND$) and split them into two equal sized groups: high and low network diversity. \tabref{tbl:topicscomparison} compares the representative keywords of the top ten topics from the topic profiles of users in these two groups. Users in high network diversity positions tend to be interested in more general topics, such as sports (``worldcup'', ``yankees'', ``lad''), current events (``ferguson'', ``oakland''), business (``profession'', ``big data''), health (``surgery'', ``obesity''), politics (``peru'', ``palestinian''), arts (``art'', ``exhibit'', ``camera''), science (``science'', ``nasa'', ``space''), promotion (``gift'', ``offer''), etc.  According to sociological theory, users in such brokerage positions spanning multiple unconnected communities are exposed to diverse information~\cite{Burt95}; therefore, it makes sense that the topics they have in common are the more general topics. On the other hand, users in positions of low network diversity focused on more specialized topics, such as hobbies (``guitar'', ``book'', ``yoga'', ``manga''),  pets (``dog'', ``cat''), family (``birthday'', ``children''), food (``oil'', ``kale''),  vacation (``journey'', ``escape'',``island''), home \& garden (``home'', ``interior'').

\subsection{Increasing Exposure to Diverse Information}

How can users increase the amount of diverse information they receive in social media? 
Do they follow more people to increase the volume of information received? Or do they move themselves into special network positions?
To examine how user effort affects information access, we split users into four classes based on the average number of tweets they post daily ($O$).
The top quartile contains the most active users, who post more than 5.3 tweets per day, the second quartile contains users who post from 3.1 to 5.3 tweets per day and the third and the bottom quartile contains from 1.9 to 3.1 and fewer than 1.9 tweets per day respectively.

\begin{figure}[t]
\begin{center}
\begin{tabular}{c}
\\
\includegraphics[width=0.9\linewidth]{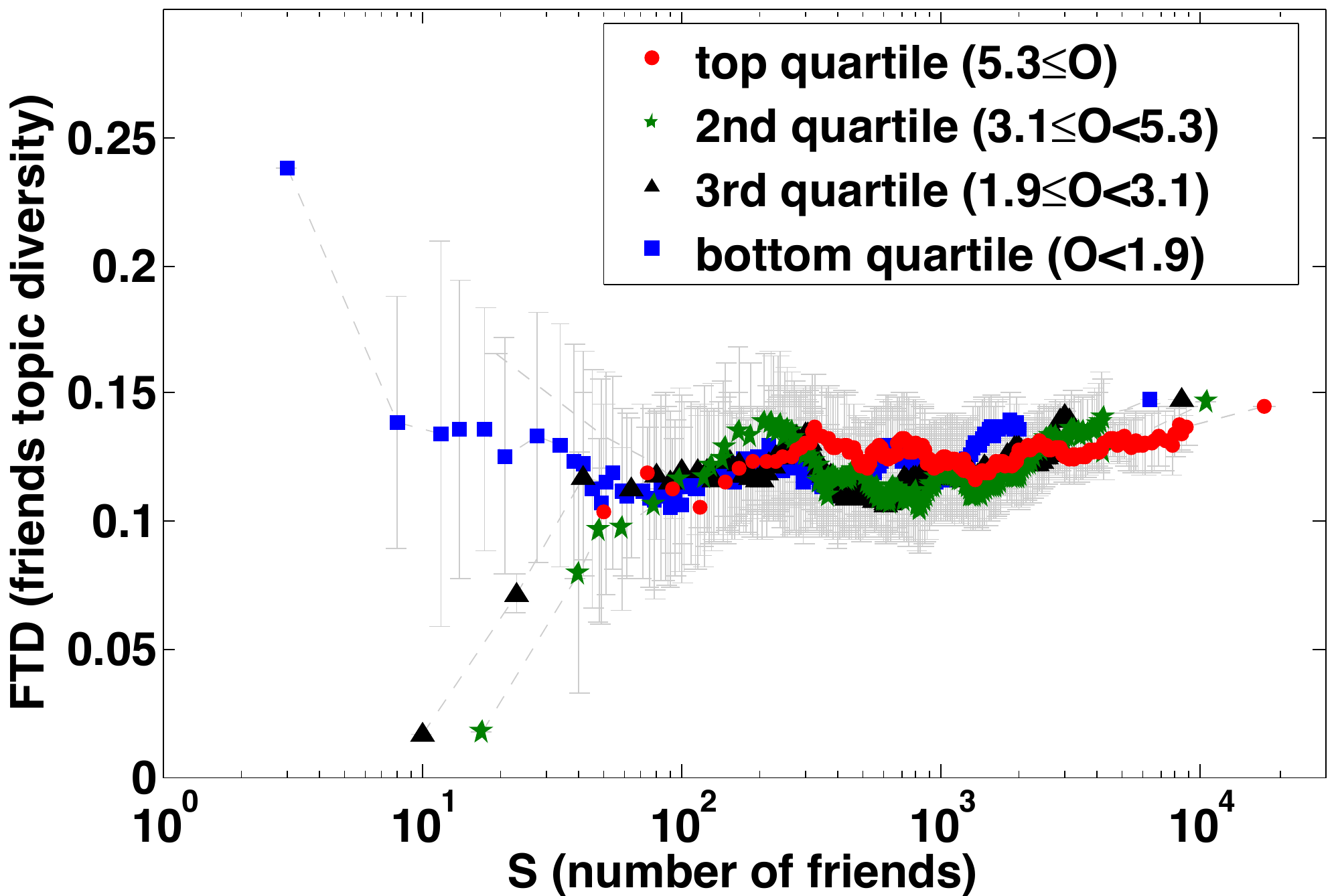}
\end{tabular}
\end{center}
\caption{
Diversity of received information as a function of user's network size. Users are divided into four populations based on their effort: red circles represent the more active users, (who post more than 5.3 tweets per day on average), green stars represent the 2nd quartile (3.1$\le$ $O_i$$<$5.3), black triangles represent 3rd quartile (1.9$\le$ $O_i$$<$3.1) and the blue squares represent that bottom quartile users (who post fewer than 1.9 tweets per day on average). We discretize values into  equal-sized bins for each quartile.
}
\label{fig:netaftdb}
\end{figure}

\figref{fig:netaftdb} shows the relationship between diversity of received information, measured by friend topic diversity ($FTD$), and user's network size ($S$), for these classes of Twitter users. The trends among these four classes of users are somewhat different, indicating that people use different strategies to access information in network. Active users who expend more effort  on Twitter (red circles in~\figref{fig:netaftdb}) increase  their exposure to diverse information by adding more friends (0.1874, p$<$.01). However, when the bottom quartile users (blue squares in~\figref{fig:netaftdb}) add friends, this actually decreases the diversity of information they are exposed to until around 100 friends. After that point, information diversity slowly increases.
For the same network size, the less active users actually receive more diverse information than the more active user until around 100 friends. Apparently, network size itself cannot provide an access to diverse information (when $S > 100$) since the network structure can vary significantly.

\begin{figure}[tbh]
\begin{center}
\begin{tabular}{c}
\includegraphics[width=0.9\linewidth]{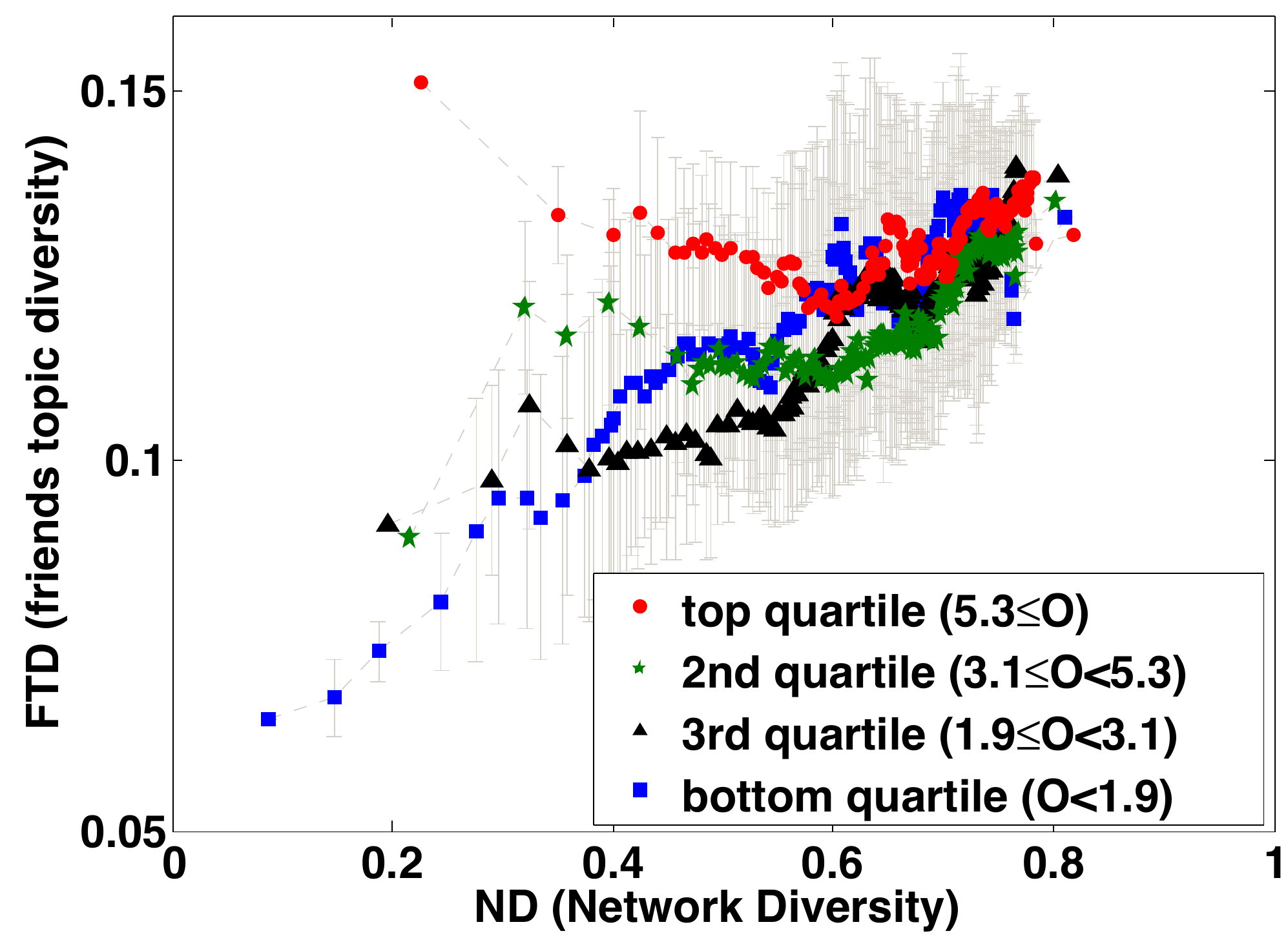}
\end{tabular}
\end{center}
\caption{
Friend topic diversity ($FTD_i$) of a user as a function of the network diversity ($ND_i$)  in the 2014 Twitter data set.
We show the average  of $FTD_i$ for the same network diversity ($ND_i$) users with their standard deviation ranges in grey color.
Users in the higher network diversity positions tend to be exposed to more diverse information, with active users receiving more diverse information regardless of their position in the network structure. We group $ND$ values into  equal-sized bins and compute the mean of both $ND$ and $FTD$ within each bin.
}
\label{fig:ndftd}
\end{figure}

In addition to network size, network position is known to play an important role in determining access to information.
In social and email communication networks, people in high network diversity positions receive more novel and diverse information~\cite{granovetter1973,Aral11,aral2012anatomy}.
We tested whether the same conclusions hold for Twitter using topics learned by the proposed model. 
Figure~\ref{fig:ndftd} shows the relationship between friend topic diversity ($FTD_i$) and structural network diversity ($ND_i$) for the four classes of users divided according to their effort.
There is a strong correlation (0.9212 (p$<$.01)) for bottom quartile users (blue squares in  Figure~\ref{fig:ndftd}), between network position and information diversity, correlation values decrease with increasing user effort (3rd quartile 0.9162  (p$<$.01) and 2nd quartile 0.7774 (p$<$.01)).
When these users place themselves in more structurally diverse position within the Twitter network, they receive on average more topically diverse tweets from friends than users who place themselves in less structurally diverse network positions. However, the correlation between $FTD$ and $ND$  for active users (red circles in Figure~\ref{fig:ndftd}) is far less, 0.3248 (p$<$.01). These users are generally exposed to more diverse information than the less active users, regardless of their network position.
Also, active users in low network diversity positions receive more diverse information than the less active users in similar positions.
These results demonstrate that the effort users are willing to invest in using social media is an important factor in access to diverse information.

\begin{figure}[t]
\begin{center}
\begin{tabular}{c}
\includegraphics[width=0.9\linewidth]{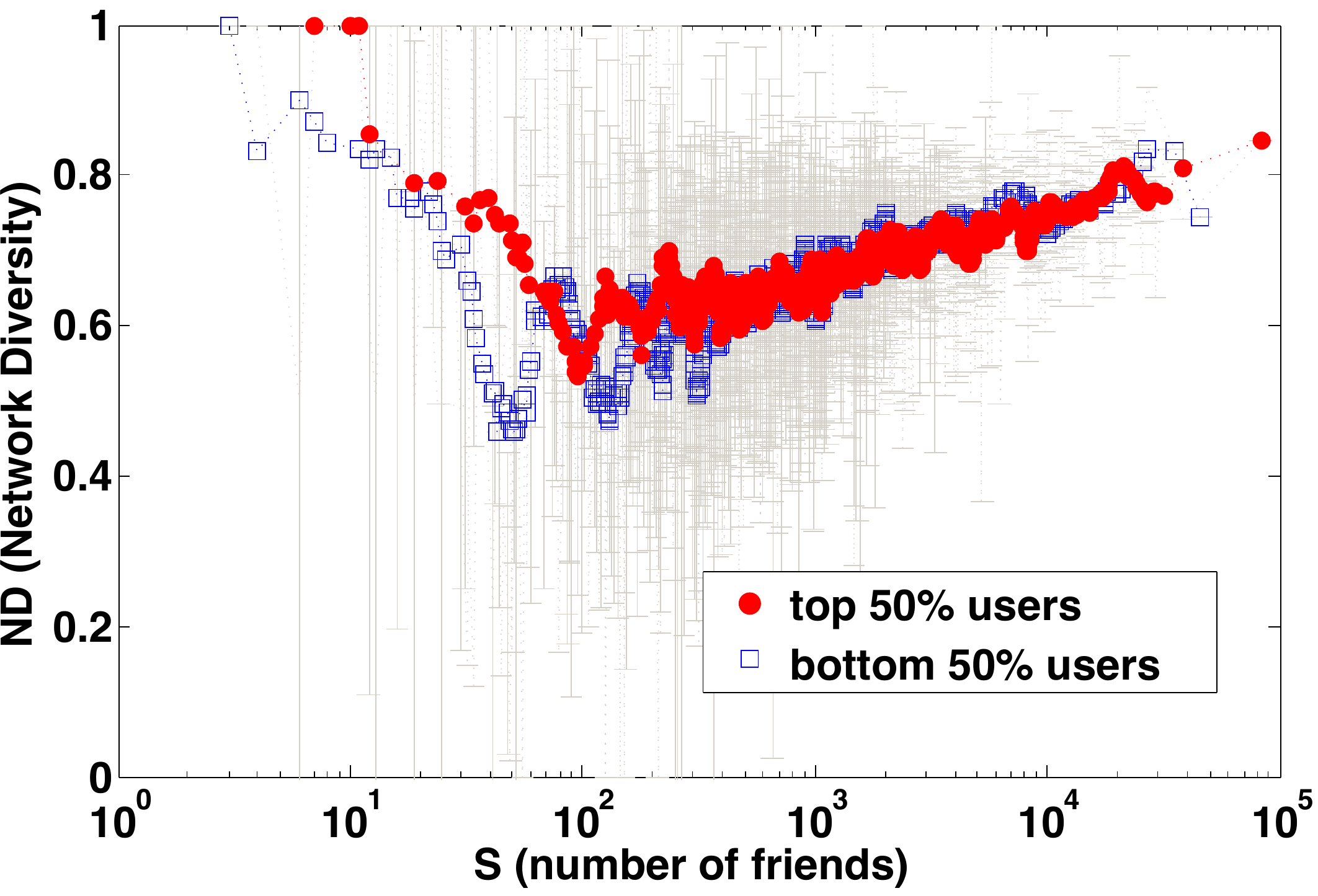}
\end{tabular}
\end{center}
\caption{
Network diversity ($ND$) as a function of the number of active friends ($S$) in the 2014 Twitter data set. We use equal-sized bins for each class.
}
\label{fig:nds}
\end{figure}

\begin{figure}[tbh]
\begin{center}
\begin{tabular}{c}
\includegraphics[width=0.9\linewidth]{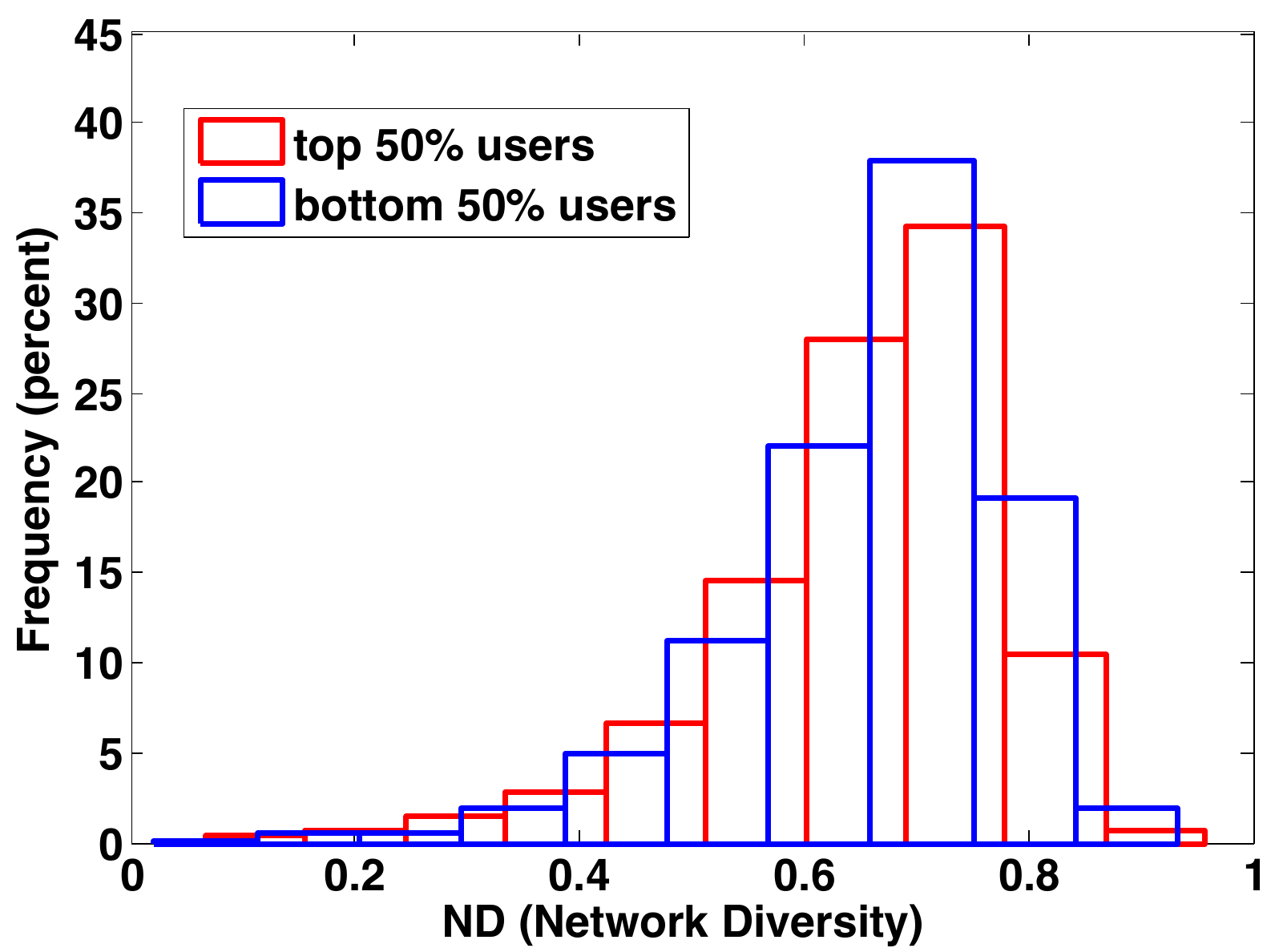}
\end{tabular}
\end{center}
\caption{
Histograms of network diversity ($ND$) of users in the 2014 Twitter data set.
Users are divided into two populations based on their effort ($O$). The peak of top 50\% users is higher than bottom 50\% users, while bottom 50\% users tend to have higher $ND$.
}
\label{fig:histfreq}
\end{figure}
Why are highly active users exposed to more diverse information? To address this question, we study how network diversity changes as users add more friends. \figref{fig:nds}  shows this relationship for users separated into two classes based on their activity or effort. Overall, network diversity increases with network size (after around 100 friends), which is not surprising since probabilistically as the number of people in a network grows, any two people are less likely to be connected to each other. Active users overall place themselves in more structurally diverse positions.

Surprisingly, network diversity initially decreases with network size for both user populations,  reaching a minimum around $S=100$. A potential explanation of this effect involves the Dunbar number. Dunbar~\cite{Dunbar} argued that finite human cognitive capacity constrains the number of social interactions individuals can manage, limiting size of social groups to about 100--200 individual. Research has validated the impact of cognitive constraints on online social interactions~\cite{goncalves2011modeling,kang2013structural}.
Similar arguments could apply to our setting. Minimum network diversity corresponds to maximal social connectivity, which in our Twitter data set occurs when users have around 100 friends. While their social networks can grow beyond that size, increasing network diversity implies that new friends are less likely to form a community.

The minimum in network diversity for the less active users occurs at lower values than for the more active users. This suggests that active users who invest more effort into using Twitter can manage larger communities of connected friends than the less-active users. This observation is in line with cognitive limits on social interactions theory: users who have a greater capacity for social interactions (or who may simply be willing to invest more time and effort in social interactions) will have more interactions on Twitter (higher activity), and they will also tend to belong to larger social groups (higher network size), simply because they are better capable of managing their social connections. At this time we cannot prove this intriguing possibility, and leave it as a question for future research.

\section{Related Work}
A pair of classic theories has linked an individual's position within a network to the novelty and diversity of information she receives through her social contacts.
The theoretical argument, known as ``the strength of weak ties''~\cite{granovetter1973}, explored the relationship between social links and the information people receive along those links. Specifically, the weak links, representing infrequent social interactions, were shown to deliver novel information to people, providing new social and economic opportunities~\cite{uzzi1997social,reagans2001networks,reagans2003network,allen2003managing}.

Burt~\cite{Burt95,Burt04,burt2005brokerage} argued that weak ties act as bridges between different communities. Individuals with many such ties are in what he termed ``brokerage positions'' in the network, which allows them with access, and benefit from, novel information residing in diverse sources.
Empirical research on mobile phone~\cite{onnela2007structure}, email communication~\cite{Aral11,iribarren2011affinity}, and online social networks~\cite{journals11074009,centola2007complex,centola2010spread}  supported the weak ties arguments about the nature of interactions on a network and its structure.

Aral \& Van Alstyne show that both structurally diverse brokerage positions in the network and high frequency communication along social ties provided access to diverse and novel information in the email communication network. In social media, Kang \& Lerman~\cite{kang2013structural} showed that increasing activity of social media friends a user follows affected how much novel information user received from them, while increasing network diversity provided access to more topically diverse information, but not the other around.
Bakshy et al.~\cite{bakshy2012role} showed that, although strong ties are individually more influential, weak ties  increased the diversity of information received.

Cognitive constraints on social interactions provide an interesting perspective on the structure and function of social networks.  
Dunbar argued that people have a limited ability, defined by their brain's capacity, to manage social interactions, which gives rise to maximum social group size~\cite{Dunbar03}. Although social media was believed to expand the size of human social networks,  research showed that the maximum number of friends that Twitter users interact with is around 100-200~\cite{goncalves2011modeling}, similar to the Dunbar number.
Cognitive constraints could also explain the findings of \cite{Aral11,aral2012anatomy}, namely that cognitive constraints create a trade-off between the complexity of social interactions (given by network diversity) and the intensity of interactions along structurally complex links, resulting in ``diversity--bandwidth trade-off.''
Unlike previous researchers, we examined how users vary in their capacity for social interactions (or activity), and how this capacity defines their level of engagement with the social media site and access to diverse information.

Recommender system~\cite{herlocker1999algorithmic,Sarwar01itembasedcollaborative,karypis2000evaluation} examines item ratings of many people to discover their preferences and recommend new items that were liked by similar people. Latent-factor models, such as probabilistic matrix factorization~\cite{salakhutdinov2008probabilistic,koren2009matrix,WangB11}, have shown promising in creating better recommendations by incorporating personal relevance into the model.
Many social recommender systems have been proposed by matrix factorization techniques for both user's social network and their item rating histories~\cite{ma2008sorec}. In addition to modeling user-item adoptions, researchers integrate social correlation between users~\cite{purushotham}, topic influences of friends~\cite{Kang13aaai}, and cognitive biases~\cite{Kang15sbp} in social recommender system.

Recommender systems often focus on understanding user preferences based on the history of observed actions to recommend possible future likes and interests.
One of the key challenge is how to increase the variety of recommended items without the expenses of sacrificing the accuracy.
The trade-off between exploration and exploitation is important to prevent over-specialization where we never recommend items outside of the history of user's actions.
Most of the current approaches focus on proposing new intra-list diversity metrics~\cite{ziegler2005improving,agrawal2009diversifying} to diversify recommendations.
Our study shows that users increase activity to access diverse information.
We can estimate how much user opens to diverse information by taking into account the engagement levels as well as the network diversity of the user.

\section{Conclusion}

The idea that network structure affects the novelty and diversity of information people receive from their social contacts has long fascinated sociologists~\cite{granovetter1973,Burt95}.
However, humans also have a finite cognitive capacity, which constraints how many social relations they are able to manage~\cite{Dunbar}.
The interplay between network structure and cognitive constraints has important implications for how people gain access to information in social networks in general, and on social media in particular. In this paper, we explored these questions using data from a popular social media platform Twitter, where users create links in order to receive information, in the form of short text messages called tweets, from other people.

One of the challenges we faced is measuring the diversity of information users receive from their friends on Twitter. We addressed this challenge by using a probabilistic model to learn users' topics of interest from the messages they receive and share on Twitter. Our model incorporates the text of messages and a user's network in a generative model of information spread. We then used learned topics to measure diversity of the information a user is exposed to as the variance of topic interests of the user's friends.

By quantifying information diversity, we can study the factors that affect information access in networks. We confirmed that network position plays an important role: users can increase the amount of diverse information they receive by increasing the structural diversity of their network position, rather than simply increasing the number of people they follow. However, we also identified user effort as an important factor mediating access to information in networks. Users who post (and consume) more messages place themselves in positions of higher network diversity than the less active users. Even when they are in structurally similar positions, the more active users receive more diverse information.
This suggests that users who invest greater effort into using Twitter may have higher cognitive capacity for processing information, or they may simply be able to devote more time to such interactions~\cite{Miritello13}.  These users curate their links so as to increase the diversity of information they receive. One mechanism for accomplishing this is to break links so as to reduce the redundancy of received information. Even when these actions do not change a user's structural position within the network, they serve to increase information diversity.
Our work underscores the importance of cognitive factors and variation in effort in access to information in networks. Work is needed to further disentangle these factors.

\section*{Acknowledgments}
This work was supported in part by AFOSR (contract FA9550-10-1-0569), by DARPA (contract W911NF-12-1-0034), and by the NSF (under grants CIF-1217605 and SMA-1360058).

\bibliography{reference}
\end{document}